\documentclass[aps,prb,amsmath,twocolumn,amssymb,floatfixng,showpacs,
superscriptaddress,footinbib,multicol]{revtex4-1}
\usepackage{graphicx}
\usepackage{epstopdf}
\usepackage{dcolumn}
\usepackage{bm}
\usepackage{color}
\usepackage{amsmath}
\usepackage{braket}
\usepackage{amsfonts}
\usepackage{soul}
\usepackage{url}
\usepackage{array,tabularx}
\usepackage{graphics}
\usepackage{times}
\usepackage{subfigure}
\usepackage[export]{adjustbox}
\usepackage[colorlinks=true,linktoc=page,linkcolor=red,citecolor=blue]{hyperref}

\begin{document}

\title{Fast wave transport in two-dimensional $\mathcal{PT}$-symmetric lattices}
\author{Sayan Jana}
%\email{sayanjana@tauex.tau.ac.il}
\affiliation{School of Mechanical Engineering, Tel Aviv University, Tel Aviv 69978, Israel}
\author{Lea Sirota}\email{leabeilkin@tauex.tau.ac.il}
\affiliation{School of Mechanical Engineering, Tel Aviv University, Tel Aviv 69978, Israel}

\begin{abstract}

We present a theoretical investigation of wave dynamics in two-dimensional non-Hermitian $\mathcal{PT}$-symmetric lattices, where onsite, as well as inter-site control couplings are employed. Our analysis shows that these couplings can be tuned to achieve a direction-sensitive group velocity enhancement beyond what is possible in the uncontrolled (Hermitian) counterpart, while ensuring that the wave packet evolution remains bounded and dynamically stable. We derive a dedicated relation between the control parameters, providing a systematic condition under which stability is guaranteed. We then study the topological properties of the non-Hermitian system at hand, and use an experimental-ready topoelecric metamaterial platform to demonstrate the non-Hermitian couplings realization, and the resulting wave dynamics. This framework paves the way to designing stable and fast wave transport in planar non-Hermitian media. 

\end{abstract}

\maketitle

%\section*{Introduction}

Parity-time ($\mathcal{PT}$) symmetry is a notable property of non-Hermitian systems -- systems in which interaction and energy exchange with the environment are permitted~\cite{ashida2020non,ding2022non}.
When the $\mathcal{PT}$-symmetric phase is restored, the eigenstates merge within the parameter space, and real eigenvalues arise even from non-self-adjoint Hamiltonians. 
This property lead to distinctive characteristics and topological structures, such as exceptional points (EPs) and rings ~\cite{bender1998real,bender2007making,zhen2015spawning}. 
As a result, remarkable features of wave dynamics were enabled and demonstrated in metamaterials. 
In particular, $\mathcal{PT}$-symmetric metamaterials have been demonstrated to support unidirectional invisibility, cloaking, focusing, coherent absorption, sensing, and more \cite{makris2008beam,lin2011unidirectional,zhu2014p,sounas2015unidirectional,fleury2015invisible,shi2016accessing,achilleos2017non,el2018non,gu2021controlling,gu2021acoustic,manna2023inner,yang2023spectral,zhao2018parity,zhu2022hybrid,li2022gain,zhu2023higher}, across both quantum and classical domains.

To realize the functionalities enabled by non-Hermiticity in general (and by $\mathcal{PT}$-symmetry in particular), the couplings within the underlying structures need typically to incorporate either onsite gain or characteristics that defy conventional classical interpretations, such as complex-valued parameters, directional interactions, or absence of restoring forces \cite{sasmal2020broadband,rosa2020dynamics,geib2021tunable,raval2021experimental,baz2022breaking,jin2022non,cui2023experimental}.
Such coupling mechanisms often necessitate the injection of external energy, often facilitated by integrating active control strategies into the system architecture. In electric circuit metamaterials this can be implemented through direct feedback devices—for instance, operational amplifiers \cite{hofmann2019chiral,zhu2023higher,halder2024circuit,jana2025invisible,jana2025harnessing}.On other platforms, such as acoustic or elastic metamaterials, the non-Hermitian couplings are usually realized by electronic controllers, which interpret and act upon real-time measurements of the system's dynamic behavior \cite{rosa2020dynamics,zhang2021acoustic,wen2023acoustic,langfeldt2023controlling,maddi2024exact,riva2025non}. 

In this work, we exploit the $\mathcal{PT}$-symmetry property for the enhancement of wave propagation velocity in two-dimensional lattice-based elastic metamaterials.
This approach has been considered in photonic lattices~\cite{szameit2011p,ramezani2012exceptional} by introducing gain and loss, combined with lattice stretching along one axis. As a result, the EPs were lifted up, leading to a steeper slope of the Dirac cones, and thus to higher group velocities. 
An equivalent model in classical one-dimensional systems has been recently proposed and realized in active topoelectrical and acoustic metamaterials \cite{PhysRevB.111.174304,wang2025supersonic}.
Therein, a dedicated relation between the control parameters was designed to keep the underlying system dynamically stable for the higher velocities. 
Motivated by that model, here we design a stable and directional group velocity enhancement mechanism in two-dimensional classical lattices, thereby facing the complexity imposed by directionality. In particular, the mechanism enables a full control of the wave propagation velocity in a preferred direction without disturbing the dynamics in other directions.
Our results are valid for any dynamics that is second order in time, where our demonstration testbed is a two-dimensional topoelectric metamaterial.

\begin{figure*}
\includegraphics[width=17 cm, valign=c]{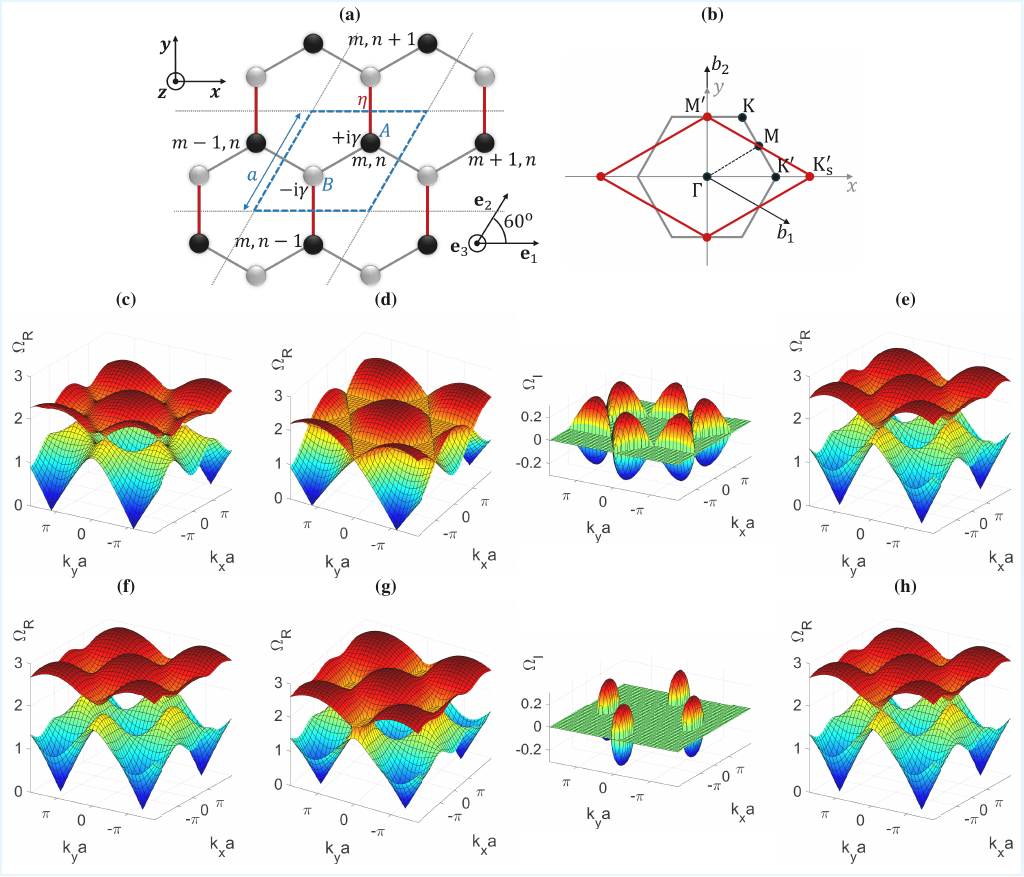}
     \caption{(a) The lattice model schematic. (b) The first Brillouin zone of the Hermitian (gray hexagon) and the non-Hermitian $\mathcal{PT}$-symmetric lattice (red parallelogram). (c)-(h) Dispersion surfaces: the nominal Hermitian case for $\eta=1$ and $\gamma=0$ (c), the unbalanced non-Hermitian case, real and imaginary ($\Omega=\Omega_R+\mathrm{i}\Omega_I$), with six triangular exceptional rings for $\eta=1$ and $\gamma=0.6$ (d), the marginal Hermitian case for $\eta=2$ and $\gamma=0$ (e), the gapped Hermitian case for $\eta=2.5$ and $\gamma=0$ (f), the unbalanced non-Hermitian case, real and imaginary, with four elliptic exceptional rings for $\eta=2.5$ and $\gamma=0.6$ (g), and the non-Hermitian $\mathcal{PT}$-symmetric case for $\eta=2.5$ and $\gamma=0.236$ (h). For (c), (e), (f), and (h) the imaginary spectrum is zero.}
    \label{fig:dispersion}
\end{figure*}

%\subsection{Model, dispersion, and the stability condition}

\begin{center}
\begin{figure*}[tb]
\includegraphics[width=11.5 cm, valign=c]{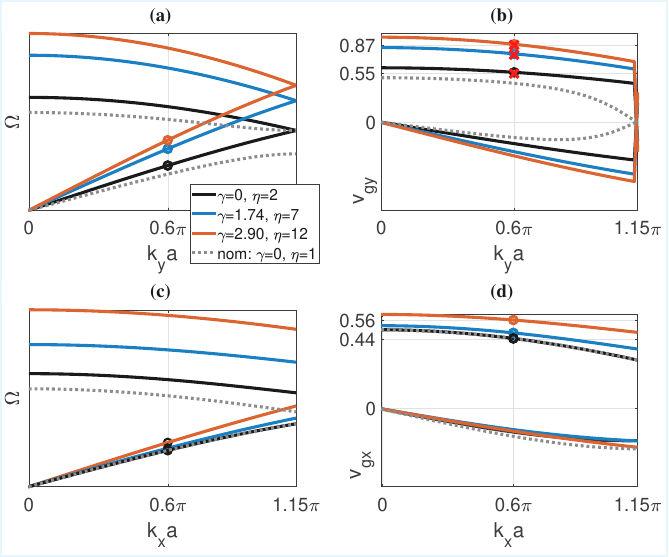} 
\caption{Group velocity increase analysis. (a), (b) The dispersion relations and the associated group velocities along the $y$ axis, for the $\mathcal{PT}$-symmetric systems with $\gamma_{pt}=0$ $(\eta=2)$, $\gamma_{pt}=1.74$ $(\eta=7)$, and $\gamma_{pt}=2.90$ $(\eta=12)$, and the corresponding $\gamma$ obtained from Eq. \eqref{eq:gamma_eta_balance}, on top of the nominal Hermitian system with $\gamma=0$ and $\eta=1$. The red crosses indicate numerical data obtained from the spacetime simulations of Fig. \ref{fig:simulations}. (c), (d) The same along the $x$ axis. The black curves in panels (a) and (b) respectively represent $\Omega_2$ and $v_{gy2}$ from Eq. \eqref{eq:vgy}.}
    \label{fig:vg}
\end{figure*}
\end{center}

%\section*{Results}

\textbf{Modeling, dispersion analysis, and group velocity enhancement.}
We consider the model in Fig. \ref{fig:dispersion}(a). The Hermitian base is a graphene-shaped lattice of the constant $a$, comprising identical ``atoms" (circles), which are connected to their nearest neighbors by identical elements (bars). The atoms and the connecting elements can represent, for example, masses and springs in a mechanical lattice, or capacitors and inductors in an electric circuit lattice (dubbed a topoelectrical metamaterial and used in Fig. \ref{fig:model_topoelectric} for dynamical demonstration). 
In the non-Hermitian setting, the lattice sites are subjected to an interlacing gain-loss pattern of strength $\gamma$ (dividing the circles into black and gray). In addition, the connections in the $y$ direction are scaled by $\eta$ (red bars). 
Assuming that the sites have a single degree of freedom $u$, the equations of motion at the $A$ and $B$ sites take the form
\begin{equation}  \label{eq:model}
\begin{split}
    \Omega_0^{-2}\Ddot{u}^A_{m,n}&=u^B_{m-1,n-1}-2 u^A_{m,n}+u^B_{m-1,n+1}\\&-\eta\left(u^A_{m,n}+u^B_{m,n}\right)+\gamma \dot{u}^A_{m,n}, \\
    \Omega_0^{-2}\Ddot{u}^B_{m,n}&=u^A_{m+1,n-1}-2 u^B_{m,n}+u^A_{m+1,n+1}\\&-\eta\left(u^A_{m,n}+u^B_{m,n}\right)-\gamma \dot{u}^B_{m,n},
\end{split}
\end{equation}
where $\Omega_0=\omega/\omega_0$, and $\omega_0$ is the characteristic frequency. 
In the dynamical equations \eqref{eq:model} we model the gain and loss effect in the transient instead of the steady-state regime, i.e. by explicitly representing it  by the on-site rates $\pm\gamma\dot{u}$ rather than by the commonly used in quantum lattices $\mathrm{i}\gamma u$ (we employ the complex-valued notation for illustration purposes only in Fig. \ref{fig:dispersion}). This turns out crucial for our approach to a stable group velocity increase.
To obtain the dispersion of the periodic system,
we substitute in Eq. \eqref{eq:model} the Bloch wave solution $\textbf{u}(t)=e^{\mathrm{i}\textbf{k}\cdot \textbf{r}}e^{-\mathrm{i}\Omega t}\overline{\textbf{u}}$, where $\overline{\textbf{u}}=[\begin{array}{cc}
   \overline{u}^A  &  \overline{u}^B
\end{array}]$ is the respective amplitudes vector, and $\textbf{r}=ma\hat{\textbf{e}}_1+na\hat{\textbf{e}}_2$ is the real space vector in terms of the lattice principal axes $\hat{\textbf{e}}_1=(1,0)$ and $\hat{\textbf{e}}_2=(\frac{1}{2},\frac{\sqrt{3}}{2})$. 
The wavevector $\textbf{k}=k_1\textbf{b}_1+k_2\textbf{b}_2$ can be then defined in terms of the inverse space axes $\textbf{b}_1=\frac{2\pi}{a}(1,-\frac{1}{\sqrt{3}})$ and $\textbf{b}_2=\frac{2\pi}{a}(0,\frac{2}{\sqrt{3}})$, as depicted in Fig. \ref{fig:dispersion}(a).
The dispersion relation is captured by the quadratic eigenvalue problem $\left|\Omega^2 \textbf{I}-\Omega \mathrm{i}\gamma\sigma_z-\mathcal{H}_0-(2+\eta)\textbf{I}\right|=\textbf{0}$, where
\begin{equation}
    \mathcal{H}_0=\left[\begin{array}{cc} 0 & f \\ f_{pt} & 0 \end{array}\right], \quad f=-\left(1+e^{\mathrm{i}k_1a}+\eta e^{\mathrm{i}k_2a}\right),
\end{equation}
and $\sigma_z$ is a Pauli matrix.  
We then define the equivalent linear eigenvalue problem $|\Omega\textbf{I}-\mathcal{H}|=\textbf{0}$ in terms of the augmented effective Hamiltonian
\begin{equation}  \label{eq:Hamiltonian}
\mathcal{H}=\left(\begin{array}{cc}
       \textbf{0}  & \textbf{I} \\
       \mathcal{H}_0+(2+\eta)\textbf{I}  & \mathrm{i}\gamma\sigma_z
    \end{array}\right).
\end{equation}
The resulting dispersion relation then takes the form
\begin{equation} \label{eq:Omega_sol}
    \Omega=\sqrt{\tfrac{1}{2}\left(2(2+\eta)-\gamma^2\right)\pm\tfrac{1}{2}\sqrt{\delta}},
\end{equation}
where 
\begin{equation} \label{eq:delta}
    \delta=\gamma^4-4(2+\eta)\gamma^2+4ff^\dagger.
\end{equation}
For the nominal Hermitian case, where $\gamma=0$ and $\eta=1$, the spectrum in Eq. \eqref{eq:Omega_sol} is purely real, as depicted in Fig. \ref{fig:dispersion}(c). The two dispersion surfaces are the standard for graphene, touching at the Dirac-like points at $\mathrm{K}=\frac{2\pi}{a}(\frac{1}{3},\frac{1}{\sqrt{3}})$, $\mathrm{K}'=\frac{2\pi}{a}(0,\frac{2}{3})$, and their counterparts by the first Brillouin zone -- the gray hexagon in Fig. \ref{fig:dispersion}(b). 
We then begin to vary $\gamma$ and $\eta$, where each of these parameters changes the dispersion surfaces in a different way. 
Keeping $\eta=1$ and increasing only $\gamma$, the system turns non-Hermitian in a broken $\mathcal{PT}$-symmetry phase, and imaginary spectrum emerges, as illustrated in Fig. \ref{fig:dispersion}(d). The exceptional rings that are formed in the merged surfaces of the real spectrum are of a triangular shape rather than the circular shape expected for the equivalent quantum system~\cite{szameit2011p,zhen2015spawning}. This is due to the quadratic form of the eigenvalue problem of the classical system in hand.
The system then becomes dynamically unstable, and the associated time response is divergent.

When $\gamma$ is kept zero, and $\eta$ is increased from 1 to 2, the system resumes Hermiticity, but the $\mathrm{K}$ point at the Brillouin zone and its mirror by the $y$ axis start to move towards each other along the $x$ axis until they are merged for $\eta=2$. 
The resulting Brillouin zone then transitions from a hexagon to a parallelogram, as illustrated in Fig. \ref{fig:dispersion}(b). 
The resulting spectrum, still purely real, is depicted in Fig. \ref{fig:dispersion}(e). The dispersion surfaces are now touching at four points instead of six, located at $\mathrm{K}'_s=\frac{2\pi}{a}(0,1)$, $M'=\frac{2\pi}{a}(0,\frac{1}{\sqrt{3}})$, and their respective reflections through the $x$ and $y$ axes. The new points are of the semi Dirac type~\cite{zhong2017semi}, which is determined by the quadratic dependency for $k_x$ and linear dependency for $k_y$.
Increasing $\eta$ further keeps the spectrum real, but creates a gap between the surfaces, as evident from Fig. \ref{fig:dispersion}(f).
If, in addition, $\gamma$ is increased above the zero, the system turns again non-Hermitian, with imaginary spectrum appearing, Fig. \ref{fig:dispersion}(g). However, the exceptional rings, centered at the four vertices of the parallelogram, are then of an elliptic shape. 

To stabilize the system while keeping it non-Hermitian, the spectrum in Eq. \eqref{eq:Omega_sol} needs to be real with nonzero $\gamma$. We thus require that $\delta$ in Eq. \eqref{eq:delta}, as well as the argument of the square root in Eq. \eqref{eq:Omega_sol}, are non-negative for all $\textbf{k}$. Therefore, $\delta$ needs to be evaluated at its minimum, which occurs for $k_1a=0$ and $k_2a=\pi$, and renders $ff^\dagger=\eta^2+4(1-\eta)$. The balancing $\gamma-\eta$ condition with minimal value of $\gamma$ that solves $\delta_{\textrm{min}}=0$ in Eq. \eqref{eq:delta}, and thus restores the $\mathcal{PT}$-symmetric phase, then reads 
\begin{equation}  \label{eq:gamma_eta_balance}
    \gamma_{pt}=\sqrt{2}\left(\sqrt{\eta}-\sqrt{2}\right).
\end{equation}
The spectrum of this non-Hermitian system is then restored to be purely real, as depicted in Fig. \ref{fig:dispersion}(h). 
The square root dependence of $\gamma$ on $\eta$ in Eq. \eqref{eq:gamma_eta_balance} is different from analogous dependence in quantum systems, for example, in photonic lattices, between the gain-loss parameter and the lattice stretching~\cite{szameit2011p,ramezani2012exceptional}. In the latter, the $\mathcal{PT}$-symmetry phase is restored for $\gamma_{pt}=\eta-2$, which is a linear dependence. In classical systems, if the steady state regime of the gain-loss is assumed, it is also possible to use a linear relation like this, but then the underlying dynamics will remain unstable even for the seemingly restored $\mathcal{PT}$-symmetry of the spectrum, as dynamical stability is a property of the transient regime~\cite{PhysRevB.111.174304}. 

%\subsection{Group velocity enhancement}

With Eq. \eqref{eq:gamma_eta_balance} at hand, we now approach the group velocity increase task. 
We calculate the dispersion relation of the $\mathcal{PT}$-symmetric case in Fig. \ref{fig:dispersion}(g) for three values of $\eta$, 2, 7, and 12, with the corresponding balancing values of $\gamma$ obtained from Eq. \eqref{eq:gamma_eta_balance}. In Fig. \ref{fig:vg}(a)-(b) and (c)-(d)
we plot the $y$ and $x$ cross-sections, respectively, of the resulting dispersion, on top of the nominal Hermitian case $\eta=1$ and $\gamma=0$, as well as the corresponding group velocities, all trimmed at $\frac{2}{\sqrt{3}}\pi$.
The highest group velocity increase is obtained along the $y$ axis, which is the $\eta$ coupling direction. 

It can be observed that the velocity increase over the Hermitian system for a given $\eta$ value is higher along the $y$ axis (Fig. \ref{fig:vg}(b)) than along the $x$ axis (Fig. \ref{fig:vg}(d)) for all the momenta. For example, at $(k_x,k_y)=(0,0.6\pi)$, for $\gamma_{pt}=1.74$ $(\eta=7)$ we have $v_{gy}=0.76$ (not ticked), which is 1.38 times higher than the Hermitian $v_{gy}=0.55$ for $\gamma_{pt}=0$ $(\eta=2)$, and 1.73 times higher than the nominal Hermitian $v_{gy}=0.44$ (not ticked). Accordingly, at $(k_x,k_y)=(0.6\pi,0)$, we have $v_{gx}=0.48$ for $\gamma_{pt}=1.74$, which is 1.09 of increase over the Hermitian $1\leq\eta\leq2$. Similarly, for $\gamma=2.90$ $(\eta=12)$, $v_{gy}$ is 1.58 times higher, with $v_{gy}=0.87$, than the velocity of the Hermitian system of $\gamma_{pt}=0$ $(\eta=2)$, and is compared to the respective $v_{gx}=0.56$. 
In particular, the increase of the group velocity in the $y$ direction in the non-Hermitian system, $v_{gy}$, compared to the Hermitian one for the stretched BZ-- the red parallelogram in Fig. \ref{fig:dispersion}(a), $v_{gy2}$, has a square root dependency on the parameter $\gamma_{pt}$, given by
\begin{equation}  \label{eq:vgy}
    v_{gy}=v_{gy2}\sqrt{\tfrac{1}{2}\gamma_{pt}+1}
\end{equation}
for all $k_y$, where $v_{gy2}=\frac{\partial \Omega_2}{\partial k_y}$, and $\Omega_2=\sqrt{\sqrt{2}}\sqrt{2}\sqrt{\sqrt{2}-\beta}$, with $\beta=\sqrt{1+\cos{\left(\frac{\sqrt{3}}{2}k_ya\right)}}$, is the Hermitian dispersion relation along $k_y$ for $\gamma_{pt}=0$ and $\eta=2$. 

\begin{figure}[tb]
    \centering 
    \includegraphics[width=6.5 cm, valign=c]{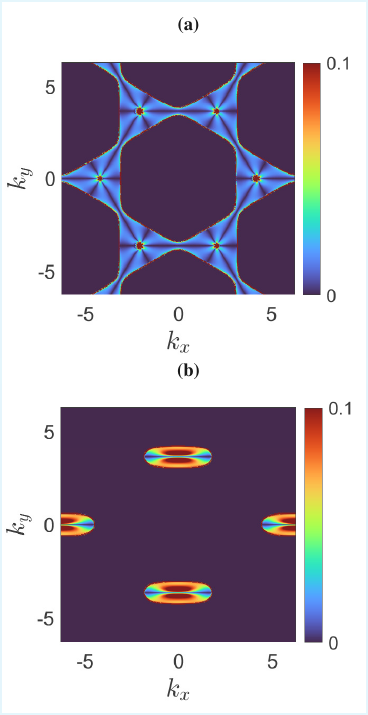}
    \caption{(a) Berry curvature distribution $F(k_x,k_y)$ according to Eq. \eqref{eq:topology_eqs}, inside and outside the exceptional surface, corresponding to the dispersions shown in Figs.~\ref{fig:dispersion}(f) and~\ref{fig:dispersion}(g) in panel (b).}
\label{fig:Berry_curvature}
\end{figure}

\bigskip

\textbf{Topological properties of the non-Hermitian system.}
The various dispersion relations in Fig. \ref{fig:dispersion} raise the interest not only for group velocity increase, but also for possible topological properties. In particular, the characterization of degeneracies in systems with complex energies relies on the topology of the complex energy dispersions. Around an exceptional point, the two levels exchange in the complex plane along a closed loop, giving rise to a half-integer vorticity \cite{keck2003unfolding,jana2023emerging} as a topological index. In contrast, in our case of various exceptional surfaces [see Figs.~\ref{fig:dispersion}(c) and (g)], the characterization is different due to their intricate structure. Inside these exceptional surfaces, the $\mathcal{PT}$ symmetry is broken, whereas it is restored outside. When a loop is taken around an exceptional surface or an isolated exceptional point [see Fig.~\ref{fig:dispersion}(h)], because the loop lies entirely in real space, the two energy levels do not exchange, resulting in a trivial zero vorticity.

To capture the underlying topology in this scenario, we examine the eigenvectors, which become meaningful due to biorthogonality. For a biorthogonal non-Hermitian system, the Berry curvature for the $n$th band is defined as
\begin{subequations}  \label{eq:topology_eqs}
\begin{align}
    \mathbf{F}_n(\mathbf{k}) &= \nabla_{\mathbf{k}} \times \mathbf{A}_n(\mathbf{k}), 
    \\ \mathbf{A}_n(\mathbf{k}) &= i \langle \tilde{u}_n(\mathbf{k}) | \nabla_{\mathbf{k}} | u_n(\mathbf{k}) \rangle,
\end{align}    
\end{subequations}
where $|u_n\rangle$ and $|\tilde{u}_n\rangle$ are the right and left eigenvectors of the Hamiltonian, respectively, which satisfy the biorthogonal normalization $\langle \tilde{u}_m | u_n \rangle = \delta_{mn}$. The Berry curvature is gauge invariant and encodes the underlying topological properties. As shown in Fig.~\ref{fig:Berry_curvature}, its distribution clearly distinguishes the unbroken and broken $\mathcal{PT}$-symmetric phases: inside the exceptional surface, where $\mathcal{PT}$ symmetry is broken, the Berry curvature $\mathbf{F}$ in Eqs. \eqref{eq:topology_eqs} is finite, while outside, where $\mathcal{PT}$ symmetry is preserved, $\mathbf{F}$ vanishes. This behavior of the Berry curvature and the associated topological distinction are unique to non-Hermitian systems and do not occur in Hermitian systems\cite{wang2024berry}. Thus, the exceptional surface acts as a boundary separating regions of zero and nonzero Berry curvature, highlighting a fundamentally non-Hermitian topology.

\begin{center}
\begin{figure*}[htpb]
\includegraphics[width=16 cm, valign=c]{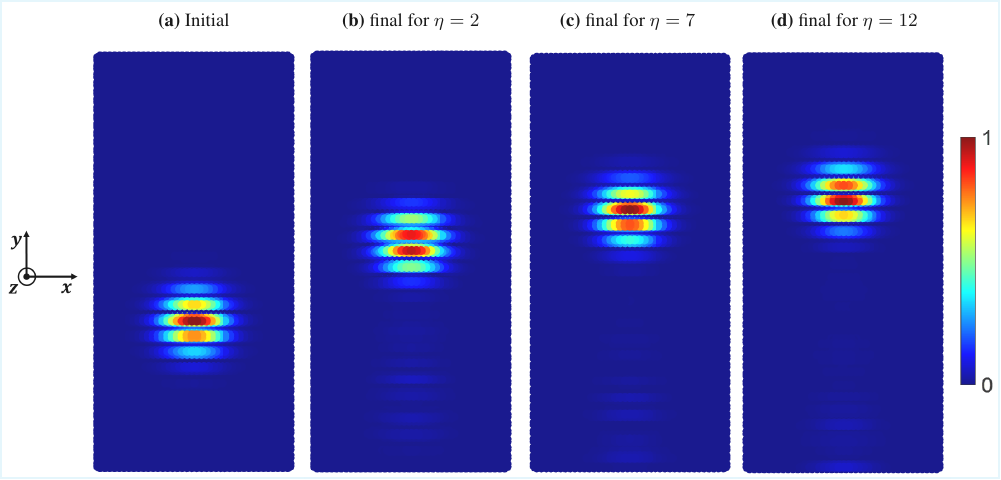}
\caption{Time domain numerical simulations and group velocity increase demonstration for the generic lattice model in Fig. \ref{fig:dispersion}(a). (a)-(d) Time domain wave packet propagation in the $y$ direction of the $\mathcal{PT}$-symmetric lattice (normalized to $\Omega_0=1$). The initial distribution (a) corresponding to $k=0.6\pi$, and the final response for the combinations $\gamma_{pt}=0$ $(\eta=2)$, $\gamma_{pt}=1.74$ $(\eta=7)$, and $\gamma_{pt}=2.90$ $(\eta=12)$, depicted in (b), (c), and (d), respectively.}
    \label{fig:simulations}
\end{figure*}
\end{center}

\textbf{Time domain realization}. 
In this section we test the time domain performance of the model derived in the previous section. We start from the generic lattice model. Then, we design an experimental-ready topoelectric metamaterial analogue, which is comprised of realistic elements and subjected to voltage excitation, and test it as well. Additionally, we test the robustness of our design to uncertainty in the electric circuit elements values.

%\subsection{Nominal model dynamics}  \label{nom}

First, we numerically simulate the time domain response along the $y$ axis of the lattice governed by Eqs. \eqref{eq:model}. By invoking the biorthogonality property of Bloch eigenvectors for the non-Hermitian system at hand, we construct an initial condition, which corresponds to the selected momentum in Fig. \ref{fig:vg}(a) for three different values of $\eta$. Complimenting it by the coordinated initial rates, and modulating by a Gaussian envelope, yields a wavepacket that propagates along the positive $y$ direction only. 
The initial wavepacket is plotted in Fig. \ref{fig:simulations}(a), and the resulting time domain responses of the lattice, at the same final time $\mathrm{t_f}$, for $\eta=2$, $7$, and $12$, are depicted in Fig. \ref{fig:simulations}(b), (c), and (d), respectively. 
The dynamical stability of the underlying non-Hermitian systems is manifested by the constant amplitudes of the responses.
The group velocities calculated from these simulations are labeled by red crosses in Fig. \ref{fig:vg}(b) on top of the analytic values, demonstrating decent alignment.

\begin{figure*}[tb]
    \centering    
    \includegraphics[width=17 cm, valign=c]{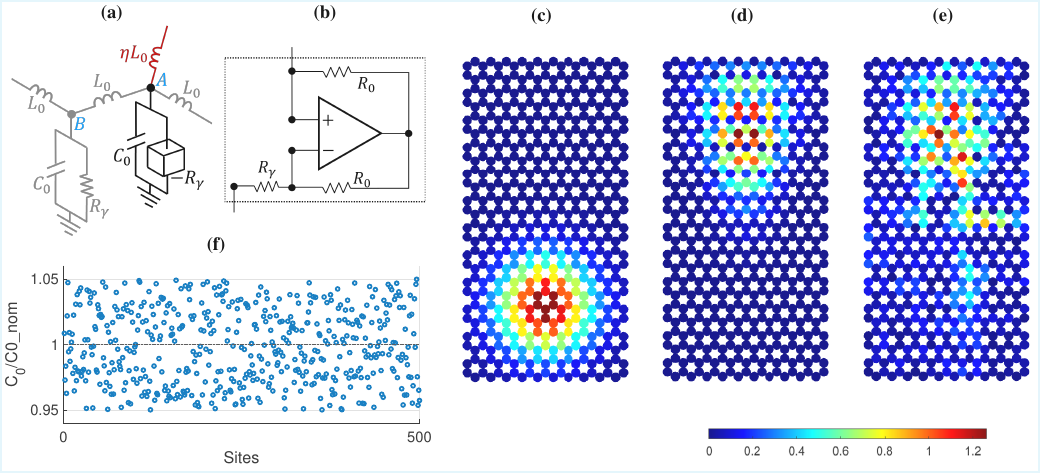}
    \caption{Topoelectrical metamaterial realization. (a) The unit cell schematics. (b) An operational amplifier in a negative impedance converting setup implementing the onsite gain element. (c) Voltage inputs location across a metamaterial of 12X10 hexagonal cells, depicted by Gaussian modulation, according to Eq. \eqref{eq:V_source} for $k=0.6\pi$. The circles represent the nodes $A$ and $B$. The color map in this panel is not to scale. (d) Snapshot of the space-time response of the non-Hermitian $\mathcal{PT}$-symmetric metamaterial for $\eta=7$ and $\gamma_{pt}$ according to Eq. \eqref{eq:gamma_eta_balance}, for nominal values of the electric circuit elements $C_0=150$ $\mathrm{nF}$ and $L_0=220$ $\mathrm{\mu H}$, and no resistance other than the controlled $R_\gamma$. (e) The same as in panel (d), but with a disorder in $C_0$, and in $R_0=23$ Ohm, which is an inherent element resistance, according to the diagram in panel (f). (f) Disorder in the values of capacitance $C_0$ and $R_0$, normalized by the nominal values. The color bar is relevant to panels (d) and (e), and indicates Volts.}
    \label{fig:model_topoelectric}
\end{figure*}

\bigskip

\textbf{Topoelectrical metamaterial model.}
To pave the way for an experimental realization, we design a realistic topoelectric metamaterial model, the unit cell of which is depicted in Fig. \ref{fig:model_topoelectric}(a). This metamaterial realizes the atoms and the connections between them of the general lattice of Fig. \ref{fig:dispersion}(a) via capacitors $C_0$ and inductors $L_0$, respectively, leading to $\Omega_0=1/\sqrt{L_0C_0}$.
To realize the loss at the $B$ we use a resistor $R_{\gamma}=\sqrt{\frac{L_0}{C_0}}\gamma^{-1}$. 
To induce a corresponding gain at the $A$ site, we implement negative resistance of $-R_{\gamma}$. This is achieved by an active system captured by a black cube, connected in parallel to the capacitor. 
It's key element is operational amplifier, Fig. \ref{fig:model_topoelectric}(b), supplied by constant voltage, and assembled in a negative impedance converting setup. Consequently, the amplifier features both forward and backward current flow through identical resistors $R_0$, where the backward flow is fed to the lattice through a $R_{\gamma}$ resistor, yielding negative effective resistance $-R_{\gamma}$ between the terminals.
The system is then governed by \eqref{eq:model}, with the onsite gain and loss $\pm\gamma$.

Next we are interested to launch a unidirectional wavepacket in the metamaterial, similarly to the lattice in Fig. \ref{fig:simulations}. However, contrary to the lattice system, in the topoelectric metamaterial we will not be able to employ initial conditions. We therefore design a pattern for external voltage sources, applied to the metamaterial nodes with a spatial modulation by the Bloch eigenvectors of the system, which correspond to the Hamiltonian in Eq. \eqref{eq:Hamiltonian}. This modulation, combined with a Gaussian envelope, generates a tailored phase shift for the voltage inputs, as
\begin{equation}   \label{eq:V_source}
    V_{n,m}(t)=
\beta_{n,m}e^{-(\mu_r+\mu_t)}
\sin\bigl( kr_{n,m} - \Omega t + \phi_{n,m} \bigr).
\end{equation}
Here, $\mu_r=(r_{n,m} - r_0)^2/4\alpha^2$ and $\mu_t=(t - r_0)^2/4\sigma_t^2$ are the spatial and temporal Gaussian modulations, with $r_0$ representing the desired distribution center and $r_{m,n}$ are the distances of the corresponding sites about this center. $\beta_{n,m}e^{\mathrm{i}\phi_{n,m}}$ is the eigenvector modulation. 
The voltage inputs locations are depicted in Fig. \ref{fig:model_topoelectric}(c). They are outlined by the Gaussian distribution component of \eqref{eq:V_source} (this is for illustration purposes, hence the values in panel (c) do not represent voltages at time zero or any other time). 
We note that the intensity distribution in Eq. \eqref{eq:V_source} is not necessary for the metamaterial excitation per se, as it can be excited even at a single site (node). 
However, a single site excitation, or an uncoordinated excitation of a few sites will cause a multidirectional propagation and a strong spatial dispersion, which will make it difficult to track the velocity increase.

The space-time response of the metamaterial to the voltage input in Eq. \eqref{eq:V_source}, for example for $\eta=7$ ($\gamma_{pt}=1.74$) and $k=0.6\pi$, is depicted in Fig. \ref{fig:model_topoelectric}(d). The maximal voltage level of the input is 1 V. The simulation is stopped before the wavepacket hits the wall. Some dispersion can be observed, as we tried to use the smallest number of sites as possible to facilitate experimental realization. It is still possible though, using the center of mass of the wavepacket, to measure the group velocity in the $y$ direction, which reads $v_{gy}/v_{gy2}=1.38$, compared to the theoretical 1.36, as expected from Eq. \eqref{eq:vgy} (the response for $\eta=2$, $\gamma_{pt}=0$ is not shown in Fig. \ref{fig:model_topoelectric}). 

%\bigskip

\textbf{Robustness to uncertainty in the metamaterial elements.}
To complete the description of the metamaterial model as an experimental-ready one, we test its performance under the assumption of uncertainty in the values of the electric circuit elements. Specifically, we simulate the metamaterial for the capacitor $C_0$ values randomly varying within $10\%$ peak to peak about the nominal value provided by the manufacturer. 
In addition, we assume that the capacitors have an inherent resistance $R_0$, varying as well within $10\%$ peak to peak about a nominal value. 

The spacetime response of the resulting metamaterial for $\eta=7$ ($\gamma_{pt}=1.74$) to the same excitation as in Fig. \ref{fig:model_topoelectric}(d) is depicted in Fig. \ref{fig:model_topoelectric}(e) (the color map indicates Volts). Although the dispersion in this case is stronger, it can be observed qualitatively that the center of mass of the final wavepacket aligns with the one in panel (d). Specifically, the measured group velocity enhancement over the one obtained for $\eta=2$, $\gamma_{pt}=0$ reads $v_{gy}/v_{gy2}=1.26$. This is quite close to the value obtained in the case of panel (d), demonstrating the robustness of the group velocity enhancement approach of Eqs. \eqref{eq:gamma_eta_balance}-\eqref{eq:vgy}. The disorder diagram is given in Fig. \ref{fig:model_topoelectric}(f).

%\section*{Discussion}
\textbf{Conclusion.}
To summarize, we proposed a model for group velocity increase in two-dimensional hexagonal lattices, based on concepts from $\mathcal{PT}$-symmetric non-Hermitian physics. Specifically, we tailored the lattice couplings by two control parameters-- gain and loss $\gamma$ at the sites, which turned the system into non-Hermitian, as well as a coupling $\eta$ between the sites in the $y$ direction only. 
With the $\gamma$ couplings alone the system featured complex-valued spectrum, indicating dynamical instability, and a consequent diverging time response. 
We found that to stabilize the system, the gain and loss contributions $\gamma$ need to be addressed in the transient regime rather than in steady-state. In the equations of motion this translated into an explicit inclusion of the onsite rates $\pm\gamma\dot{u}$ rather than the complex-valued terms $\pm\mathrm{i}\gamma u$ that are commonly used in quantum systems.
This approach enabled us to derive the dedicated nonlinear relation between $\gamma$ and $\eta$ in Eq. \eqref{eq:gamma_eta_balance}, which eliminated the imaginary spectrum by restoring the $\mathcal{PT}$-symmetry phase, thus stabilizing the system, and in the same time supporting a controlled increase of the group velocity. 

The highest increase, compared to the Hermitian case, was achieved in the $y$ direction, and featured a square root dependence on $\gamma$, given by Eq. \eqref{eq:vgy}. 
The corresponding group velocity increase in the $x$ axis direction is significantly smaller, as can be seen in Fig. 4 of the revised manuscript. This property enables one to fully control the wave propagation velocity in a preferred direction without disturbing the dynamics in other directions, implying a direction-sensitive controlled group velocity enhancement in 2D systems.

The topology analysis of the underlying two-dimensional non-Hermitian system revealed that inside the exceptional surface, where $\mathcal{PT}$ symmetry is broken, the Berry curvature $\mathbf{F}$ from Eqs.~\eqref{eq:topology_eqs} remains finite. In contrast, outside the surface, where $\mathcal{PT}$ symmetry is preserved, $\mathbf{F}$ vanishes. Such a contrast in Berry curvature, and the resulting topological distinction, is a hallmark of non-Hermitian systems and has no counterpart in Hermitian systems.

A numerical experiment in the time domain confirmed the theoretical predictions. In particular, an equivalent topoelectric metamaterial model was designed. It is composed of interconnected inductors and capacitors forming the Hermitian base layer, whereas the non-Hermitian control parameters were realized by resistors for the loss, and active feedback amplifier circuits for the gain, as illustrated in Fig. \ref{fig:model_topoelectric}. This experimental-ready model is based on the dynamics of realistic electric circuit elements, and is driven by an external voltage source using a signal generator. Using a coordinated space-dependent and modulated by non-Hermitian eigenvectors phase shift between the actuated metamaterial sites, we managed to induce a unidirectional propagation of the wavepacket, similarly to the generic model driven by initial conditions.

In addition, this model enables an explicit account for imperfections, such as losses due to electrical resistance, and/or disorder due to variation in the nominal values of capacitance, inductance, or resistance, provided by the manufacturer. In the uncertainty analysis we found that despite an overall $10\%$ of variation in the elements values, as well as losses that are inherent to the circuit and are not part of the gain-loss balance, a substantial group velocity increase was still observed.
Our model can be further utilized for a fully-controllable directional fast transport of waves on other platforms, such as elastic or acoustic metamaterials.

%\section*{Data Availability}

%The data is available from the authors upon a reasonable request.

%....................
\section*{Acknowledgements}
%................

\textit{This research was supported in part by the Israel Science Foundation Grants No. 2177/23 and 2876/23.}

%\section*{Author Contributions}

%L. S. initiated and supervised the project, and performed the dispersion and group velocity analysis. S. J. performed the time domain analysis, the topology analysis, and the uncertainty analysis. Both authors wrote the manuscript and contributed to the discussion.

%\section*{Competing Interests}

%The authors declare no competing interests.

%-----------------------------------------
%\appendix

%\renewcommand{\thefigure}{A\arabic{figure}}
%\renewcommand{\theequation}{A\arabic{equation}}
%\setcounter{equation}{0}
%\setcounter{figure}{0}
%-----------------------------------------
%~~~~~~~~~~~~~~~~~~~~~~~~~~~~~~~~~~~~~~~~~~~~~~~~~~~~~~~~~~~~~~

%\section{Dispersion relation and $\mathcal{PT}$-symmetry}  \label{dispersion_details}

%\bibliographystyle{IEEEtran}

\bibliographystyle{naturemag}

\bibliography{paper}

\end{document}